\documentstyle[12pt,qqa4lart,psfig]{article}
\input tcilatex

\QQQ{Language}{
American English
}

\begin{document}

\author{P. N. Bhat }
\date{Tata Institute of Fundamental Research\\Homi Bhabha Road, Mumbai 400 005,
India.}
\title{Pachmarhi Array of \v Cerenkov Telescopes\thanks{%
A talk delivered at the Tata Institute of Fundamental Research during
``Perspectives in High Energy Astronomy \& Astrophysics'', an International
Colloquium to commemorate the Golden Jubilee year of TIFR (August 12-17,
1996).}}
\maketitle

\begin{abstract}
This talk is based on the Very High Energy Gamma Ray Astronomy observations
planned to be carried out at Pachmarhi in the central Indian state of Madhya
Pradesh using the well known atmospheric \v Cerenkov technique. The
development of a ground based array of 25 \v Cerenkov telescopes is
currently underway at Pachmarhi situated at an altitutde of about a
kilometer. Using this array it is proposed to sample the \v Cerenkov light
pool at various distances from the shower core in order to estimate the
lateral distribution parameters of the shower. Simulation studies have shown
that these parametrs would enable one to distinguish gamma ray initiated
showers from those by cosmic ray charged particles, thus significantly
improving the signal to noise ratio. After summarizing the genesis of VHE
gamma ray astronomy in our institute we will discuss the scientific
motivation of this concept of enriching the gamma ray signal as compared to
the standard imaging technique. The current status of the detector
development and the expected results will be presented.
\end{abstract}

\section{Introduction\ }

The gamma ray region of the electromagnetic spectrum has been the last to be
successfully exploited as a channel for astronomical investigation. Although
it was realized as far back as 1958 (Morrison, 1958), the detection
techniques were difficult and the fluxes were low. The chief motivation for
the study of $\gamma $-rays is that they are produced in the high energy
particle interactions. The close association between cosmic rays and $\gamma 
$-rays makes $\gamma $-ray astronomy a special branch of cosmic ray
astrophysics. A careful study of the intensity spectrum of $\gamma $-rays
and their time variability may reveal the physical conditions and
acceleration mechanisms of cosmic rays with in the source.

Before $\sim $1980,``Gamma Ray Astronomy'' was interpreted as a branch of
Space Science and presumed to terminate where satellites ceased to be
useful, i.e. $\sim $1 GeV. Today the situation is quite different. Very high
energy gamma ray astronomy, covering the energy range 10$^2$-10$^4$ GeV, has
generated lot of interest as many new groups all over the world are entering
the field. Further the field has come of age and one can say with certainty
that some VHE $\gamma $-ray sources do exist.

\section{Genesis of $\gamma $-Ray Astronomy in TIFR}

At the Tata Institute of Fundamental Research we have a long history in High
Energy Physics. The interest in air \v Cerenkov radiation started in the
early 60's when we were engaged in the study of characteristics of high
energy nuclear interactions in the tens of GeV range - using cosmic rays at
mountain altitudes before the first CERN accelerator. The experimental set
up at the Cosmic Ray Laboratory at Ootacamund in the Southern Indian State
of Tamilnadu consisted of a triple arrangement comprising of an air \v
Cerenkov counter, a multiplate cloud chamber and a total absorption
spectrometer one below the other. The hadron interaction was recorded in the
cloud chamber, its total energy was measured by the TASS which was
essentially a total ionization calorimeter and the air \v Cerenkov counter
above acted as a threshold counter and served the important purpose of
distinguishing between pions and protons in the energy range 10-40 GeV. The
\v Cerenkov counter was 4 m high and the light was collected by a parabolic
search light mirror of 1m diameter.

Immediately after the discovery of Pulsars, a series of experiments were
designed in 1969 to search for very high energy $\gamma $-rays from Pulsars
using the atmospheric \v Cerenkov technique (Chatterjee {\it et al}., 1970,
71). The first atmospheric \v Cerenkov telescope consisted of two parabolic
mirrors and the activity continued with a progressively increasing
sensitivity. The collection area of mirrors were progressively increased to
6.4 m$^2$ (using 10 mirrors of 0.9 m diameter) in 1977 to 20 m$^2$ in 1979,
by adding 8 more mirrors of 1.5 m diameter. The mirrors were deployed in the
form of a compact array and the signals from neighboring mirrors were added
linearly to form a bank. A time coincidence between the banks was used as
the trigger. Later majority logic between different banks was used to
generate an event trigger.

\section{Past land marks}

\subsection{Compact Array:}

\subsubsection{Crab Pulsar Observations}

During the 1976-77 observing season the Crab pulsar was observed for 27
hours using 10 mirrors of 0.9 m diameter with a mask of 1$^{\circ }$ FWHM.
The event arrival times were converted to the arrival times at the solar
system barycenter. Using the contemporaneous radio ephimeris for Crab, the
pulsar phasogram is derived. Two peaks of 3.6$\sigma $ and 2.2$\sigma $
significance with a separation of 0.42$\pm $0.03 as seen in the MeV-GeV data
were observed. Figure 1 shows the phasogram of Crab pulsar in the TeV energy
range. The estimated steady flux of VHE $\gamma $-rays of energy above 4.5
TeV is estimated to be 2$\times $10$^{-12}$ photons cm$^{-2}$s$^{-1}$. Ever
since no observations (before applying any cuts) carried out at Ooty or
Pachmarhi showed any evidence for pulsed steady emission from this pulsar
(Gupta {\it et al.}, 1978).

\begin{figure}
\centerline{\psfig{figure=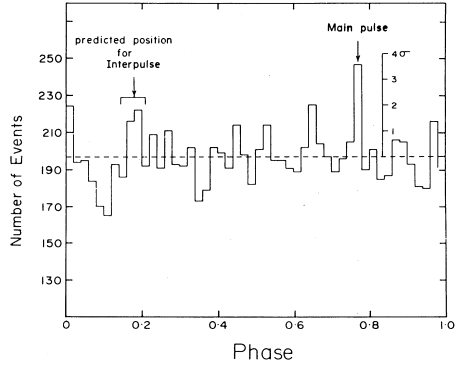,height=5.0cm,angle=0}}
\caption{Phasogram of PSR 0531+21 for VHE $\gamma$-rays of energy $>$ 4.5 TeV. Two signal peaks separated by the characetristic 0.42 of the pulsar phase could be seen.}
\label{Figure 1:}
\end{figure}
\subsubsection{Vela Observations}

Vela pulsar (PSR 0833-45) is a gamma ray pulsar which is the brightest $%
\gamma $-ray source in the GeV range. Despite the large angle this source
could just be observed from Ooty. This source was observed from Ooty for the
first time in February, 1979. The observations were continued in
February-March, 1979. The data from 33 hours of observation showed two
distinct peaks in the phasogram as shown in figure 2. The integral flux of
Gamma rays of energy $>$ 14 TeV from this source is estimated to be (5$\pm $%
1) 10$^{-13}$ photons cm$^{-2}$s$^{-1}$. Because of inadequate time keeping
the absolute pulsar phase could not be established (Bhat {\it et al.}, 1980).

\begin{figure}
\centerline{\psfig{figure=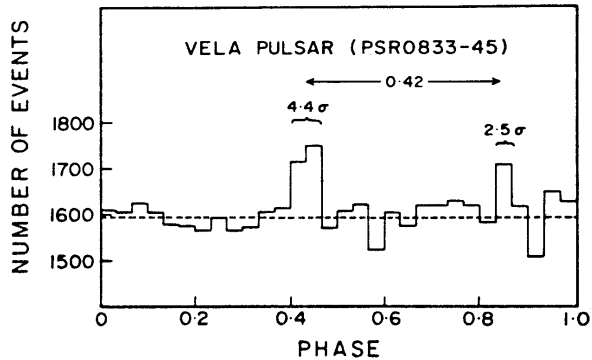,height=5cm,angle=0}}
\caption{Phasogram of about 5000 showers from the direction of Vela Pulsar (PSR 0833-45).  The data were collected during February-March 1979.}
\label{Figure 2:}
\end{figure}
\subsection{Distributed Array:}

The above observations were carried out at the back yard of the Cosmic Ray
Laboratory where the \v Cerenkov telescopes were initially deployed in the
form of a compact array. After a series of tests involving the timing of \v
Cerenkov front we showed that it is possible to determine the arrival
direction of the shower accurately by the well known method of
triangulation. This would allow us to improve the signal to noise ratio by
being able to reject off-axis showers which are due to cosmic rays. Hence it
was decided to build a distributed array of \v Cerenkov telescopes in order
to make use of this idea in 1980. Because of paucity of space we had to move
and build the array at place about 11 km away. The new array consisted of 8
large mirrors at the center of a circle of radius 55 m in the form of a
compact cluster while 10 small mirrors were situated symmetrically on the
circumference. While central cluster was used for generating the trigger the
outer mirrors were used for timing information. The uncertainty in the
arrival direction estimate is $\sim $ 0.3$^{\circ }$ if $\geq $ 6 timing
telescopes are triggered in an event. Most of events however had only about
3 timing mirrors triggered resulting in the angular accuracy of 0.6-0.7$%
^{\circ }$ for vertical incidence. One of the important drawbacks of a
distributed array is that since the mirrors are distributed the energy
threshold is invariably higher compared to that for a compact array. The
remaining 2 small mirrors were operated at the old site (Gupta, 1983).

An added advantage of the distributed array is that the number of triggered
timing telescopes is an indication of the energy of the primary thus
enabling one to provide an energy spectrum of gamma ray signal from a
celestial source. This is a unique capability available for the first time
in the history of the field of TeV $\gamma $ -ray astronomy.

\subsection{Steady signal from Veal Pulsar}

Observations on this source continued almost every year following the
initial success. The data taken during 1979-80, 1982-83 and 1984-85 with
differing exposues and energy thresholds (9.3, 4.7 \& 5.4 TeV respectively)
were independently analyzed. The phasograms generated using contemporaneous
radio ephimeris of this pulsar showed a small but persistent peak at the
same phase bin over the years which increased in strength when a cut is
applied to the data to reject 50 \% of the higher energy events based on the
pulse height data (see figure 3). The combined a phasogram showed an
impressive 4.0 $\sigma $ signal whose position coincided with the main
optical pulse of the pulsar\ (Bhat {\it et al}., 1987).

\begin{figure}
\centerline{\psfig{figure=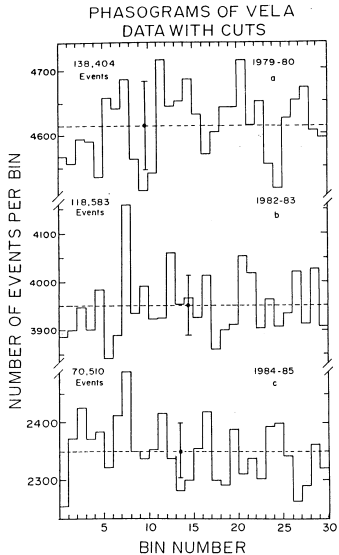,height=7.5cm,angle=0}}
\caption{Phasogram of the 3 datasets for Vela Pulsar after applying cuts to reject higher energy showers.}
\label{Figure 3:}
\end{figure}
Since the different data segments had different energy thresholds one could
derive an energy spectrum of VHE $\gamma $-rays for this source for the
first time. The slope of the integral energy spectrum in this energy range
is estimated to be $-(2.5\pm 0.3)$.

\subsection{A GRB from Crab Pulsar?}

On the other hand the Crab pulsar did not show any evidence for steady
emission. However there was an interesting detection of a transient emission
lasting for about 15 mins from this source on January 23, 1985 at 1711 hrs
UT from this source. The phasogram (shown in figure 4) in the TeV energy range
showed a strong ( 5.1 $\sigma $) peak that coincided with the radio main
pulse amounting to a gamma ray flux of (2.5$\pm $0.6)10$^{-10}$ photons cm$%
^{-2}$s$^{-1}$ at E$_\gamma $ $>$ 1.2 TeV (Bhat {\it et al}., 1986). This is
the second detection of flaring activity of a similar time scale and
strength in Crab pulsar in the TeV range, the first being in 1982 detected
by the Durham group (Gibson {\it et al.}, 1982).

\begin{figure}
\centerline{\psfig{figure=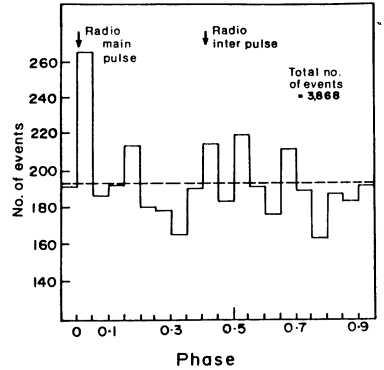,height=5cm,angle=0}}
\caption{Phasogram of PSR 0531+21 in TeV $\gamma$-rays during the 15 min 1711-1726 UT on January 23, 1985.}
\label{Figure 4:}
\end{figure}
\subsection{TeV $\gamma $-rays from Geminga}

Geminga (PSR 0633+23) is the strongest $\gamma $-rays source in the sky in
the medium energies. A periodicity of 237 ms was seen in the MeV $\gamma $%
-ray signal from this source (Hermsen{\it \ et al.}, 1992). This source was
observed from Ooty during 1984-85 using the distributed array amounting a
total of 30 hrs. exposure. When searched for a periodic signal in this
dataset a modest signal consisting of two peaks separated by 0.5 of the
pulsar phase was seen, thus establishing the first ever detection of TeV $%
\gamma $-rays from this source. The peak positions coincided with those seen
at MeV energies. The estimated integral flux of TeV $\gamma $-rays above 0.8
TeV is (2.1$\pm $ 0.8) 10$^{-11}$ photons cm$^{-2\text{ }}$s$^{-1}$(see
Vishwanath {\it et al}., 1993a).

\section{High Energy Gamma Ray Observatory (HEGRO)}

Since the inception a persistent effort was made to increase the sensitivity
of the $\gamma $-ray telescopes by (a) improving the hardware deployed in
the observations, like using better and faster data recording system \&
increasing the mirror area and (b) by developing more techniques to help us
reject the background events more efficiently like measuring the direction
of the primaries as well as measuring pulse heights which are proportional
to the \v Cerenkov light produced by them. There is yet another parameter
which can improve the sensitivity further viz. the exposure time. Ooty
weather being what it is the total exposure one could get on a source was
severely limited by poor observing conditions. Hence it was soon decided to
move the entire array to a new site at Pachmarhi ( longitude: $78^{\circ
}~26^{\prime }~E$, latitude: $22^{\circ }~28^{\prime }~N$ \& altitude: 1075
m) in the central Indian state of Madhya Pradesh. After a systematic study
of meteorological data of this place was chosen which is expected to allow
much longer exposures on sources of our interest. All the mirrors were
deployed in the form of a compact array initially to achieve as low an
energy threshold as possible for $\gamma $-rays. The observations were
resumed within 2 months of shifting.

\subsection{A TeV $\gamma $-ray burst from Her X-1}

The first source being observed from this site was the Her X-1 which by then
has been accepted to be a source of TeV $\gamma $-rays. During a run on
April 1986, a large increase in event trigger rate was observed lasting for
a period of about 14 minutes and no such increase was observed in the chance
coincidence rate which was also being monitored in real time. This is by far
the largest busrt (42 $\sigma $) ever seen in this energy range from any
source(see figure 5). During the bursting period the time averaged $\gamma $%
-ray flux was estimated to be 1.8 $\times $10$^{-8}$ photons cm$^{-2}$ s$%
^{-1}$ for E$_\gamma $ $>$ 0.4 TeV (Vishwanath {\it et al.}, 1989). This
source has been known to be active during certain epochs during its 35 day
cycle observed from X-ray observations.

\begin{figure}
\centerline{\psfig{figure=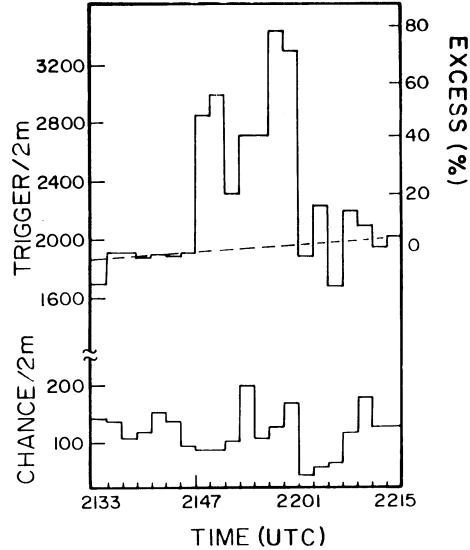,height=7.5cm,angle=0}}
\caption{The trigger rate and the chance coincidence rate for every 2 minutes for Her X-1. The right hand scale shows the percentage excess over the cosmic ray background.}
\label{Figure 5:}
\end{figure}
\subsection{Discovery of PSR 0355 + 54}

The motivation to observe this short period (156 ms) radio pulsar was its
extremely low timing noise which might indicate the presence of an outer gap
in the pulsar magnetoshpere near the light cylinder. The presence of an
outer gap may mean the existence of energetic particles which in turn could
potentially produce TeV $\gamma $-rays (Cheng et al., 1988). In addition,
two giant glitches were detected from this pulsar in 1986 (Lyne, 1987). From
our observations of the Vela pulsar which shows frequent glitches, there
seems to be an increased $\gamma $-ray activity following a glitch. Hence we
observed this source using a compact array during the end of 1987. During an
observation period of $\sim $ 25 hours a convincing evidence of a steady
emission of TeV $\gamma $-rays was detected from this source for the first
time. The phasogram, shown in figure 6, derived using contemporaneous pulsar
elements showed a 4.3 $\sigma $ peak at a phase of 0.53 with respect to the
radio pulse amounting to a $\gamma $-ray flux of 7.9 x 10$^{-12}$ photons cm$%
^{-2}$ s$^{-1}$ at E$_\gamma >$ 1.3 TeV (Bhat {\it et al.}, 1990)

\begin{figure}
\centerline{\psfig{figure=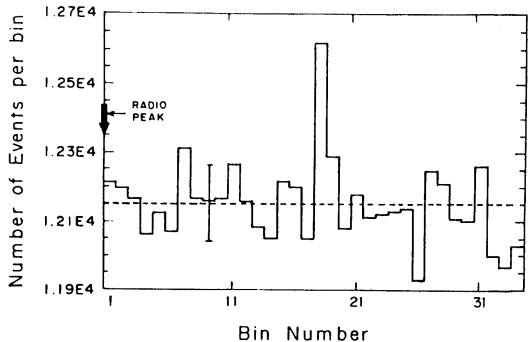,height=5cm,angle=0}}
\caption{$\gamma$-ray phasogram for 428,639 events from the direction of PSR 0355+54 showing a 4.3 $\sigma$ peak at a phase of 0.53, with respect to the radio pulse.}
\label{Figure 6:}
\end{figure}
\subsection{Distributed \v Cerenkov Telescopes}

In September 1988, the Pachmarhi array was modified in order to observe
transient emissions. The available hardware was separated into 4 groups
A,B,C \& D and were deployed at the corners of a rectangle 80 m x 90 m. Each
group consisted of 2 large (1.5 m diameter) mirrors and 2 smaller (0.9 m
diameter) parabolic mirrors. Transient emission, if any, could be
independently detected by these telescopes. Two of them could be monitoring
the off-source region during each observation. At the center of the
rectangle was a fifth array (E) consisting of 4 smaller reflectors in order
to study the energy threshold dependence of the episodal emission.

\subsubsection{A second Crab burst}

This source was observed continuously during 1988-90 period using this
modified array. On January 2, 1989 a 5 minute transient emission was seen
from this source by the 5 telescopes independently all of which were
observing the source. The signal strengths were consistent with the
different $\gamma $-ray energy thresholds. This transient emission from the
Crab pulsar, with a chance probability of 9$\times $10$^{-4}$, was the
second observed by our group. The time averaged flux during the 5 minutes
was 1.4 $\times $ 10$^{-10}$ photons cm$^{-2}$ s$^{-1}$ for E$_\gamma >$ 2.8
TeV. The pulsar phasogram during this episode showed a strong 6 $\sigma $
peak coinciding with the radio main pulse as shown in figure 7. Some
activity in the inter-pulse region also could be seen in the phasogram
(Acharya {\it et al.}, 1992).

\begin{figure}
\centerline{\psfig{figure=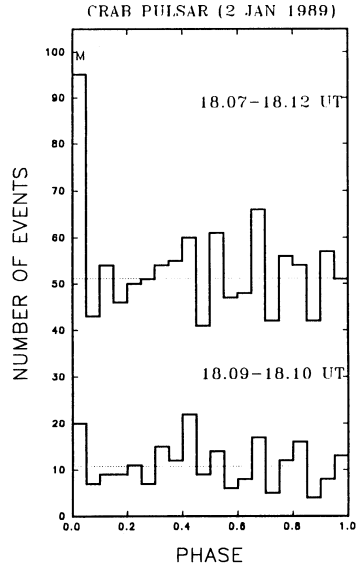,height=7.5cm,angle=0}}
\caption{Top: Phasogram of PSR 0531+21 during the 5 min burst. Bottom: Phasogram for the 1 min interval during the burst. M, I refer to the radio main and inter-pulse positions respectively.}
\label{Figure 7:}
\end{figure}
In addition to the TeV $\gamma $-ray sources mentioned above several other
sources have also been observed both from Ooty and Pachmarhi which however
did not show any evidence for the emission of TeV $\gamma $-rays. They are
the nearest pulsar PSR 0950$+$08, the short period X-ray binary 4U0115$+$63,
the millisecond pulsar PSR 1957$+$20 etc.

\section{Pachmarhi Array of \v Cerenkov Telescopes}

\subsection{Polemics}

In a constant effort to improve the signal to noise ratio further, it was
realized that one has to reduce the background due to hadronic cosmic ray
showers. Monte-Carlo simulations have shown that differences between hadron
induced and $\gamma $-ray initiated cascades could be exploited to reject
cosmic ray initiated showers (Hillas and Patterson 1987, Rao and Sinha 1988,
Knapp and Heck, 1995). The simulations have also shown that electromagnetic
cascades are flatter in the lateral distribution of \v Cerenkov photons and
more compact in the angular size of the \v Cerenkov images than those
initiated by cosmic ray primaries.

Several groups have used some of the above parameters to reject the cosmic
ray background and enhance the signal (Weekes 1989, Baillon {\it et al.,}%
1994, T\"umer {\it et al.,}1985 and Goret et al., 1993). The Whipple group
has detected steady emission of TeV $\gamma $-rays from Crab nebula using \v
Cerenkov imaging technique (Vacanti et al., 1991 and Punch et al.,1992) and
have established this source as a `standard candle' of TeV $\gamma -$rays.

\begin{figure}
\centerline{\psfig{figure=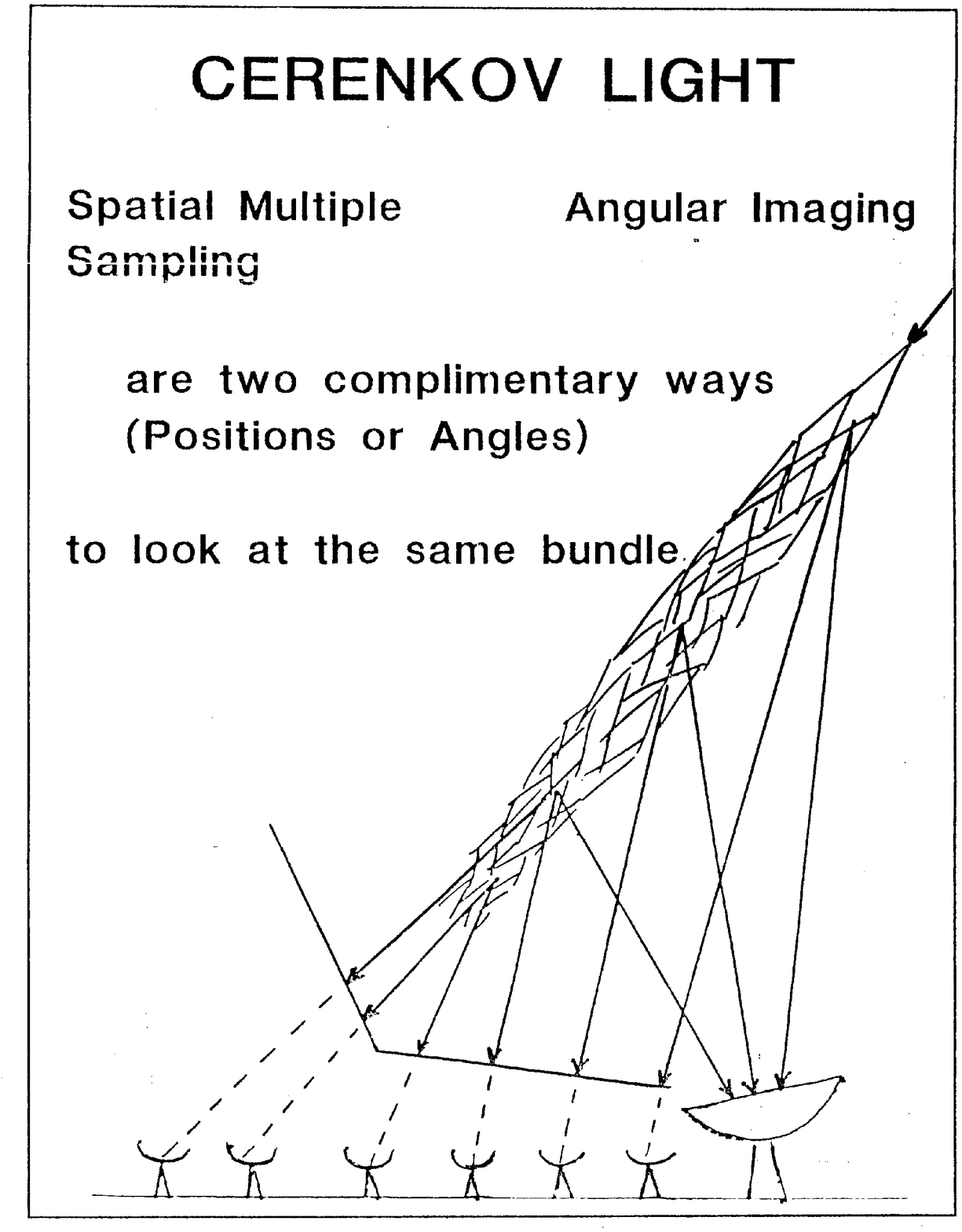,height=7.5cm,angle=0}}
\caption{}
\label{Figure 8:}
\end{figure}
However, not enough attention has been given to the lateral distribution
aspect of the atmospheric \v Cerenkov radiation to reject cosmic ray
initiated background. As shown in figure 8, angular imaging and spatial
multiple sampling are two complimentary ways to examine the same bundle.
However one looks at different types of distinguishing features between $%
\gamma $-ray and cosmic ray primaries in the two techniques. The lateral
distribution of \v Cerenkov photons have a ``hump'' at distances of 120-140
m from the shower core in $\gamma $-ray initiated cascades only (Rao and
Sinha 1988). Further, the shower to shower fluctuations in the lateral
distribution of \v Cerenkov photons is much less for a $\gamma $-ray
initiated shower compared to cosmic ray initiated showers, which are rather
`bumpy' due to the contribution of \v Cerenkov light from muons. The average
lateral distribution of \v Cerenkov photon densities, as calculated by
Hillas and Patterson ( Hillas and Patterson 1990), clearly show a flat
distribution up to the ``hump'' region for $\gamma $-ray initiated showers
and a steeper distribution for proton initiated ones. Rao and Sinha (Rao and
Sinha 1988) have shown that the signal to noise ratio could at least be
improved twice by exploiting these very differences. Calculations made for
various array configurations at Pachmarhi showed (Vishwanath et al.,1993)
that unambiguous identification of $\gamma $-ray initiated showers could be
made with a large number of detectors sampling the \v Cerenkov photons.
Accordingly, the set-up at Pachmarhi is being modified and augmented.

\subsection{The array}

We have decided to pursue the latter technique viz. to measure the lateral
distribution parameters in order to distinguish the photon primaries from
the more abundant hadronic primaries. The entire available area of $%
85~m~\times ~100~m$ at HEGRO Pachmarhi has been filled with an array of 25
\v Cerenkov telescopes. Each telescope consists of 7 parabolic (f/d $\sim $
1) mirrors mounted symmetrically making up a total reflector area of $%
4.4~m^2 $. These parabolic reflectors are fabricated locally and
their optical quality is such that a point source image is $<$ 1$^{\circ }$.
The reflectors are equatorially mounted and independently steerable both in
E-W and N-S directions. The reflector orientation and tracking are
controlled by a computer automated system (Gothe, {\it et al}., 1997) to an
accuracy of $\pm 0^{\circ }.1$. A fast photomultiplier (EMI 9807B) is
mounted at the focal plane of each reflector behind a $3^{\circ }$ diameter
mask.

The existing large and small mirrors will be deployed in the form of a
compact array at the center. These will generate an independent trigger and
hence will work like a compact array within the larger array, having a
significantly lower energy threshold. From this one will be able to derive
an energy spectrum of gamma rays from a source

Event arrival times are derived from a Global Position Satellite (GPS)
receiver having an absolute time keeping accuracy of $\pm 100~ns$. The pulse
heights and the times of arrival of pulses at each PMT will be recorded
using LeCroy ADC and TDC modules respectively. A new distributed data
acquisition system based on several PC 486's networked together using a
LINUX system is currently being developed (Bhat, 1996). Event triggers will
be generated by each of the 25 telescopes and tagged for identification. The
idea of a distributed data acquisition is to minimize the loss of
information due to transmission through long coaxial cables. Special cables
(RG 213) of shorter lengths will be used for analog pulse transmission from
the phototubes to the nearest signal processing center. Event triggers will
be generated from the royal sum of all the analog pulses from all the 7
mirrors in a telescope. The measurement of time differences of triggers from
various telescopes will enable us to measure the arrival direction correct
to 0.2-0.3$^{\circ }$.

Each event data will consist of the event arrival time to an accuracy of $%
\pm 1~{\mu }s$, amplitude and relative time of arrival of pulses at each
bank, trigger information and other relevant house-keeping informations etc
will be recorded at each of the 4 field signal processing centers while the
relative arrival times between the telescopes will be recorded at the
central data acquisition system. The data will be recorded in the internal
hard disks which will be collated off-line.

A Computer Automated Rate Adjustment and Monitoring System (CARAMS) is
developed for setting the high voltages on individual phototubes such that a
preset rate of pulses above the set threshold of 30 mV is generated (Bhat,
1996a). This will also monitor the counting rates from all the phototubes
through out the run. Similarly, the optimum count rate from each photo tube
is estimated using the count rate variation as a function of the
discriminator threshold. A software package (Automated Rate Measurement At
different Discriminator thresholds, ARMADA) to derive this curve is also
developed. This will run prior to CARAMS before starting an observation.

\subsection{Photon \& Energy Thresholds}

If we use a the royal sum of the analog pulses from the 7 mirrors in a
telescope to generate a trigger, it has been estimated that the trigger rate
will be around 1000/min. The night sky background flux at Pachmarhi was
measured to be 1.2 $\times $ $10^8$ photons cm$^{-2}$ s$^{-1}$ (Bhat and
Mehta, 1996). Using this it is estimated that the threshold photon density
is $\sim 32\ \gamma $ m$^{-2}$. Using the estimated photon densities through
monte carlo simulations at various $\gamma $-ray energies (Ong, 1995), this
photon density threshold translates to an energy threshold of $\sim $ 350
GeV for $\gamma $-rays.

\subsection{Technical Progress}

Eight of the 25 telescopes are already in position and the rest will be
erected in about 3-4 month's time.

The new distributed data recording system as well as networking between PC's
carrying out signal processing in the field signal processing centers (FSPC)
are being developed. Each of the FSPC's will have independent clocks which
will be synchronized with the GPS through the network. The FSPC's are
constantly monitored and controlled from the central station. Also the data
recording system in the central station communicates with the PC's which
control the high voltages on the individual phototubes as well as the PC
which controls and monitors the movement of the telescopes.

We expect to see the first light around October, 1997.

\section{References}

\begin{itemize}
\begin{enumerate}
\item  Acharya, B. S., Bhat, P. N., Gandhi, V. N., Ramana Murthy, P. V.,
Sathyanarayana G. P. and Vishwanath, P. R., 1992, {\it Astron. \& Astrophys}%
. {\bf 258}, 412.

\item  Baillon, P., {\it et al.}, 1994, {\it Astroparticle Phys}., {\bf 1,}
341.

\item  Bhat, P. N., Gopalakrishna, N. V., Gupta, S. K., Ramana Murthy, P.
V., Sreekantan, B. V., Tonwar, S. C. and Vishwanath, P. R., 1980, {\it %
Astron. \& Astrophys.,} {\bf 81}, L3.

\item  Bhat, P. N., Ramana Murthy, P. V., Sreekantan, B. V.and Vishwanath,
P. R., 1986, {\it Nature}, 319, 127.

\item  Bhat, P. N., Gupta, S. K., Ramana Murthy, P. V., Sreekantan, B. V.,
Tonwar, S. C. and Vishwanath, P. R., 1987, {\it Astron. \& Astrophys.}, {\bf %
178}, 242

\item  Bhat, P. N., Acharya, B. S., Gandhi, V. N., Ramana Murthy, P. V.,
Sathyanarayana, G. P. and Vishwanath, P. R., 1990, {\it Astron. \&
Astrophys.,} {\bf 236}, L1

\item  Bhat, P. N., 1996, HECR Technical Note.

\item  Bhat, P. N., 1996a, HECR Technical Note.

\item  Bhat P. N. and Mehta, S., 1996, HECR Technical Note.

\item  Chattejee, B. K. {\it et al}., 1970, Nature, {\bf 225}, 839.

\item  Chattejee, B. K. {\it et al.}, 1971, Nature, {\bf 231}, 126.

\item  Cheng, K. S., Alpar, M. A., Pines, D. and Shaham, J., 1988, {\it %
Astrophys. J.}, {\bf 330}, 835.

\item  Gibson, I. A. {\it et al.}, 1982, {\it Nature}, {\bf 296}, 833.

\item  Goret, P. {\it et al}., 1993, {\it Astron. \& Astrophys.,} {\bf 270},
401.

\item  Gothe, K. S. {\it et al}., 1997, {\it An Automated \v Cerenkov
Telescope Orientation System}, (under preparation).

\item  Gupta, S. K., Ramana Murthy, P. V., Sreekantan, B. V. and Tonwar, S.
C., 1978, {\it Astrophys. J.}, {\bf 221}, 268.

\item  Gupta, S. K., 1983, {\it Ph. D. Thesis}, University of Bombay
(unpublished).

\item  Hermsen, W., Swanenburg, B. N., Buchheri, R., Scarsi, L. and Sacco,
B., {\it et al.}, 1992,{\it \ IAU Circular No. 5532}.

\item  Hillas, A. M. and Petterson, J. R., 1987, {\it Very High Energy Gamma
Ray Astronomy} (Ed: K. E. Turver), Reidel Publishing Co., 243.

\item  Hillas, A. M. and Petterson, J. R., 1990, {\it J. Phys.,} {\bf G 16},
1271.

\item  Knapp, J. and Heck, D., 1995, Extensive {\it Air Shower Simulation
with CORSIKA: A User's Manual}, KfK Report.

\item  Lyne, A. G., 1987,{\it \ Nature}, {\bf 326}, 569.

\item  Ong, R. A., 1995, {\it Towards Major Atmospheric \v Cerenkov
Detector-IV,} Padova Ed: M. Cresti, p 261.

\item  Morrison, P., 1958, {\it Nuovo Cim.},{\bf \ 7}, 858.

\item  Punch, M., {\it et al}., 1992, {\it Nature,}{\bf \ 358}, 477.

\item  Rao, M. V. S. and Sinha, S., 1988, {\it J. Phys}., {\bf G 14}, 811.

\item  T\"umer, O. T., {\it et al}., 1985, Proc. XIX ICRC, La Jolla, {\bf 1}%
, 139.

\item  Vacanti, G., {\it et al}., 1991, {\it Astrophys. J.}, {\bf 377}, 467.

\item  Vishwanath, P. R., Bhat, P. N., Ramana Murthy, P. V. and Sreekantan,
B. V., 1989, {\it Astrophys. J.}, {\bf 342}, 489.

\item  Vishwanath, P. R. {\it et al.}, 1993, {\it Towards Major Atmospheric
Cerenkov Detectors - II}, Ed: R. C. Lamb, Calgary, 115.

\item  Vishwanath, P. R, Sathyanarayana, G. P., Ramana Murthy, P. V. and
Bhat, P. N., 1993a, {\it Astron. \& Astrophys.,} {\bf 267}, L5.

\item  Weekes, T. C., 1989, {\it Astrophys. J.}, {\bf 342}, 379.$\frac {}{}$
\end{enumerate}
\end{itemize}

\end{document}